\documentstyle[psfig]{lamuphys}
\makeatletter
\let\chapter\hid@chapter
\makeatother

\begin{document}
\pagenumbering{arabic}

\title{Spectra of High Redshift Galaxies
using a Cluster as a Gravitational Telescope}

\author{D.\,Mehlert\inst{1},
S.\,Seitz\inst{2},
R.\,P.\,Saglia\inst{2},
T.\,L.\,Hoffmann\inst{2},
I.\,Appenzeller\inst{1,3},
R.\,Bender\inst{2},
U.\,Hopp\inst{2},
R.-P.\,Kudritzki\inst{2},
A.\,W.\,A.\,Pauldrach\inst{2}}

\institute{Landessternwarte Heidelberg,
K\"onigstuhl,
D-69117 Heidelberg,
Germany
\and
Universit\"atssternwarte M\"unchen,
Scheinerstra\ss e 1,
D-81679 M\"unchen,
Germany
\and
Max-Planck-Institute f\"ur Astronomie,
K\"onigstuhl,
D-69117 Heidelberg,
Germany}

\authorrunning{Mehlert et al.}
\titlerunning{Spectra of High Redshift Galaxies
through a Gravitational Telescope}
\maketitle


\vspace{-0.3cm}
\section{1E\,0657 and the Gravitational Telescope Idea}

Tucker et al.\ (1998) identified the extended Einstein source 1E\,0657
with a cluster of galaxies at a redshift of $z = 0.296$,
showing a velocity dispersion, measured from 13 galaxies,
of $\sigma_{\rm gal} \approx 1200\,{\rm km/s}$.
X-ray data from ROSAT and ASCA indicate that merging of at least two
subclusters occurs in this highly luminous cluster.
Tucker et al.\ (1998) derive a temperature of the hot gas in
1E\,0657 $kT \approx 17\,{\rm ke\!V}$\footnote{
Yaqoob (1998) reanalyzed the X-ray data and derived a temperature
of only $kT \approx 12\,{\rm ke\!V}$, which, however, still makes it
one of the hottest clusters known yet.},
which makes it the hottest cluster known so far.
The total mass of 1E\,0657 within 1 Mpc, derived from the X-ray data,
is about $2 \times 10^{15} M_{\sun}$.

\begin{figure}
\centerline{\hbox{\psfig{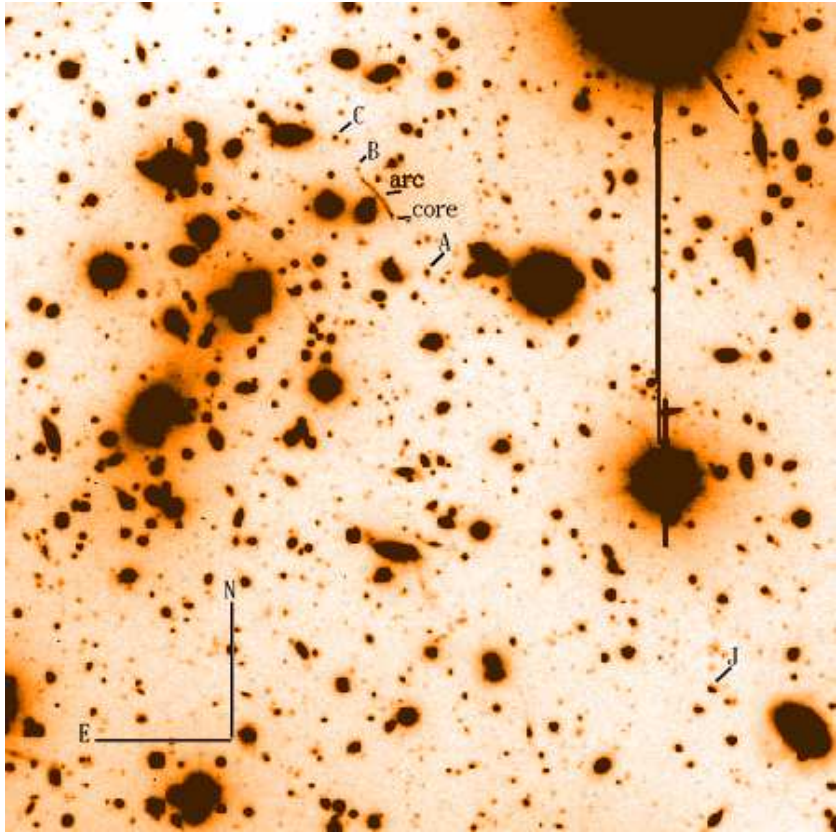}}}
\caption{R-band image of the galaxy cluster 1E\,0657 taken during
the commissioning of FORS in December 1998
(field of view (FOV) = $3.2' \times 3.2'$;
$0.2''$ per pixel;
exposure time = $3 \times 600$\,s;
seeing = $0.65''\dots 0.8''$).
The positions of the arc and some other objects (see Section 3) are indicated.
North on the top, East to the left.}
\label{fig:cluster}
\end{figure}

Most important for our project is the fact that optical images
of 1E\,0657 show the presence of a large gravitational arc
$\approx 1'$ NW of the main cluster's center.
Figure~\ref{fig:cluster}
and \ref{fig:spec_arc} shows part of an R-band image obtained with the
Focal Reducer Spectrograph (FORS) at the
Very Large Telescope (VLT)
during the FORS commissioning time in December 1998 to test the
imaging properties of the instrument (Appenzeller et al.\ 1999).
The gravitational arc itself is visible above and to the left of
the image center.
It has a length of $13''$ and a width $\la 0.8''$, which is
of the order of the seeing FWHM.
We distinguish a pure ``arc'' of $10''$ length,
as the part of the source extremely stretched by the lensing effect,
and a $3''$ ``core'' at its SW end, which lies outside the caustic
and is mapped into one fairly undistorted but still magnified image.
Using a simple lens model for the cluster, we derived a magnification of
the original background sources (probably situated at high redshift)
of $\approx 15\dots 17$ in flux.
Hence the galaxy cluster 1E\,0657 acts as a very efficient
gravitational telescope
(Fort, Mellier, Dantel-Fort 1997),
brightening distant background sources by up to 3 mag.

\section{Spectroscopic Observations}

Since the arc in the images of 1E\,0657 was suspected to be a
galaxy at high redshift, magnified by the galaxy cluster,
we obtained long-slit spectra of the arc (see Figure~\ref{fig:cluster})
with FORS in December 1998 to determine its redshift.
Using the grism 150I and a slit width of $1''$ we covered the
spectral range from $\lambda = 3300\dots 9200$\,\AA\ with a
spectral scale of 5.6\,\AA/pixel.
In total we gathered 1.5\,h of exposure time with a seeing of
about $0.5''$ and reached a signal-to-noise (s/n) of 7 for
the $10''$ of the arc itself and 5 for the $3''$ of the core image
at its SW end.
Moreover, we obtained the spectra of some other interesting objects
that also happened to fall on the slit by chance coincidence.

Standard reduction (bias subtraction, correction for flat field variation,
cosmic ray elimination) as well as rebinning to the observed wavelength was
applied to all obtained spectra, using MIDAS routines.

\section{Spectra and Preliminary Results}

\begin{figure}[b]
\centerline{\hbox{\psfig{figure=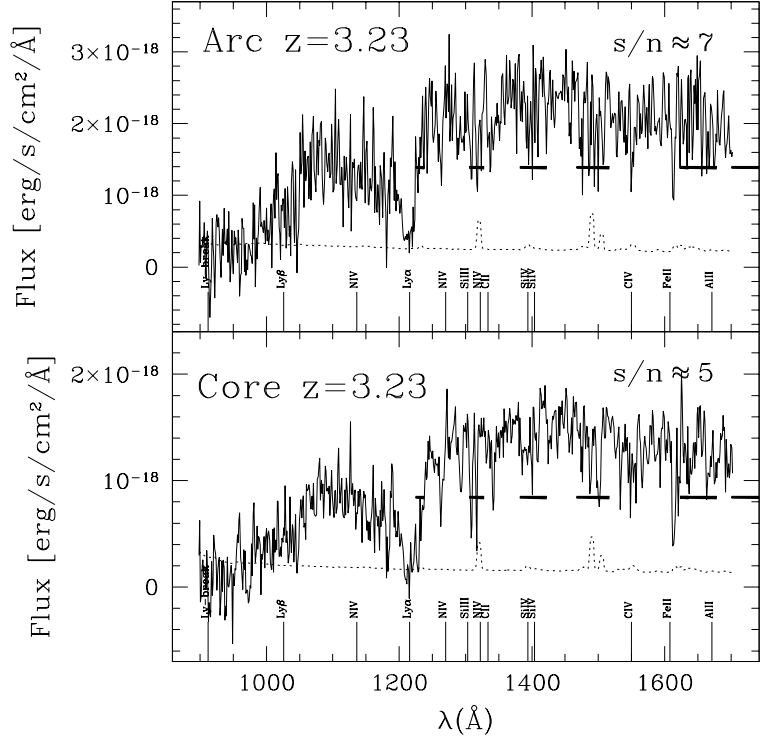,width=10cm}}}
\caption{Flux calibrated spectra of the arc ($10''$, top panel)
and the core ($3''$, lower panel).
The Ly$_{\alpha}$ absorption line as well as the position of some
expected metal absorption lines are indicated.
The horizontal bars indicate the position of prominent night sky blends
(blueshifted to the rest frame of the galaxy).
The dotted lines indicate the size of the $1\sigma$ error level of the flux.}
\label{fig:spec_arc}
\end{figure}

Figure~\ref{fig:spec_arc} shows the rest frame spectra of the
``arc'' and the ``core'' image.
The most prominent feature in both spectra is the Ly$_{\alpha}$
absorption line from which we derived a redshift of $z = 3.23$.
It is obvious that the two spectra have, in fact,
their origin in the same background source, being rich in hot stars.
From the ratios of some of the metallic lines indicated in
Figure~\ref{fig:spec_arc} we estimate that the
metallicity of the dominating stellar population should be low.

\begin{figure}
\centerline{\hbox{\psfig{figure=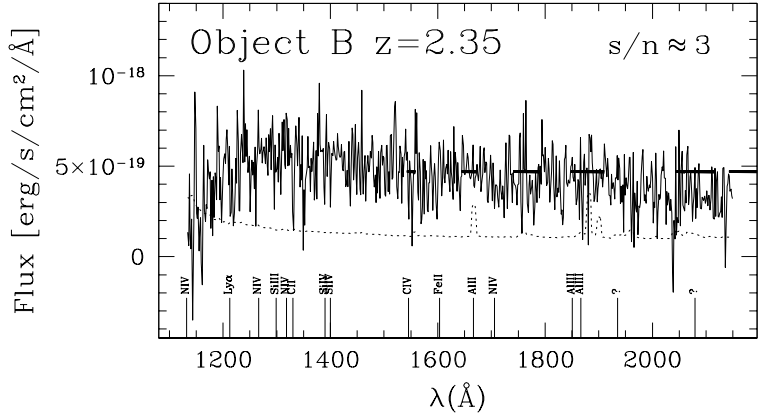,width=10cm}}}
\caption{Spectrum of Object~B, which lies a few arcsec NE of the arc.
Ly$_{\alpha}$ absorption and metal lines are indicated as in
Figure~\ref{fig:spec_arc}.
The horizontal bars and dotted lines have the same meaning as in
Figure~\ref{fig:spec_arc}.}
\label{fig:spec_b}
\end{figure}

Figure~\ref{fig:spec_b} shows the spectrum of object~B,
which lies only a few arcsec NE of the arc.
It is obvious that object~B does not belong to the arc system.
From this fact and the two-dimensional flux distribution of the arc
we conclude that the arc is a so-called ``cusp arc'',
similar to the case of A370 (Fort, Mellier 1994).
Since the s/n of object~B's spectrum is only about 3,
the estimated redshift is rather uncertain.

\begin{figure}
\centerline{\hbox{\psfig{figure=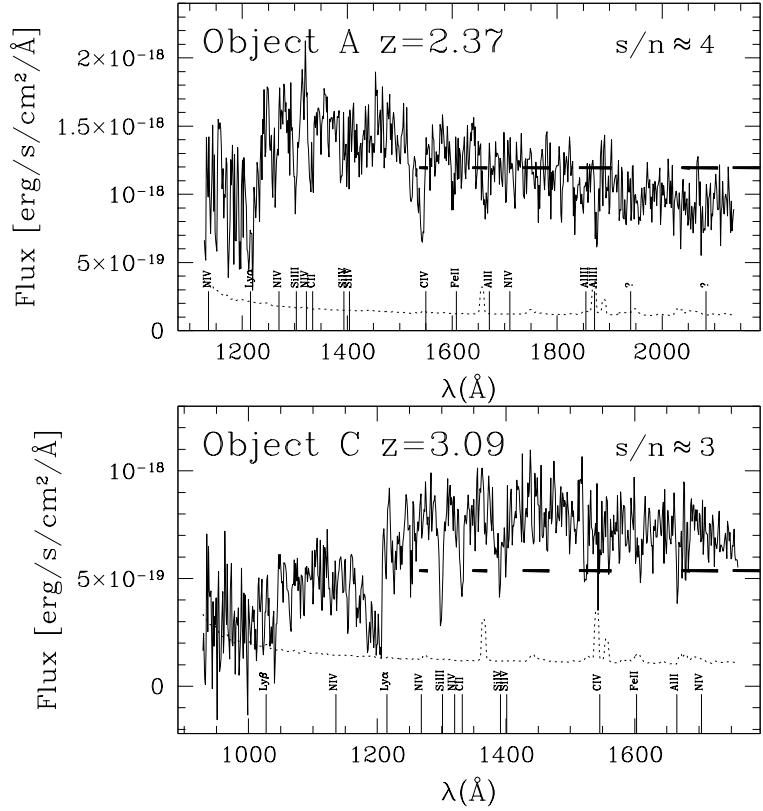,width=10cm}}}
\caption{Spectra of objects A (top panel) and C (lower panel),
which lie $\approx 20''$ SW and $\approx 10''$ NE of the arc, respectively.
Ly$_{\alpha}$ absorption and metal lines are indicated as in
Figure~\ref{fig:spec_arc}.}
\label{fig:spec_ac}
\end{figure}

Due to the lens magnification we expect that near the critical line
(i.e., the arc), galaxies have higher redshifts than galaxies of the
same brightness in the field (for example, Franx et al.\ 1997).
In fact, objects A ($\approx 20''$ SW of the arc)
and C ($\approx 10''$ NE of the arc) also turned out to be
young distant galaxies (Figure~\ref{fig:spec_ac}).
Again the most dominant feature is the Ly$_{\alpha}$ absorption line,
which leads to redshifts of $z = 2.37$ and $3.09$, respectively.
From the continuum slope redwards of Ly$_{\alpha}$ we estimate the
temperature of the hottest O~stars in Object~A to be about
$T \approx 25000\,{\rm K}$, strongly depending on the presence of
UV extinction due to intrinsic dust.
In object~C Ly$_{\alpha}$ may have a weak emission component.

\begin{figure}
\centerline{\hbox{\psfig{figure=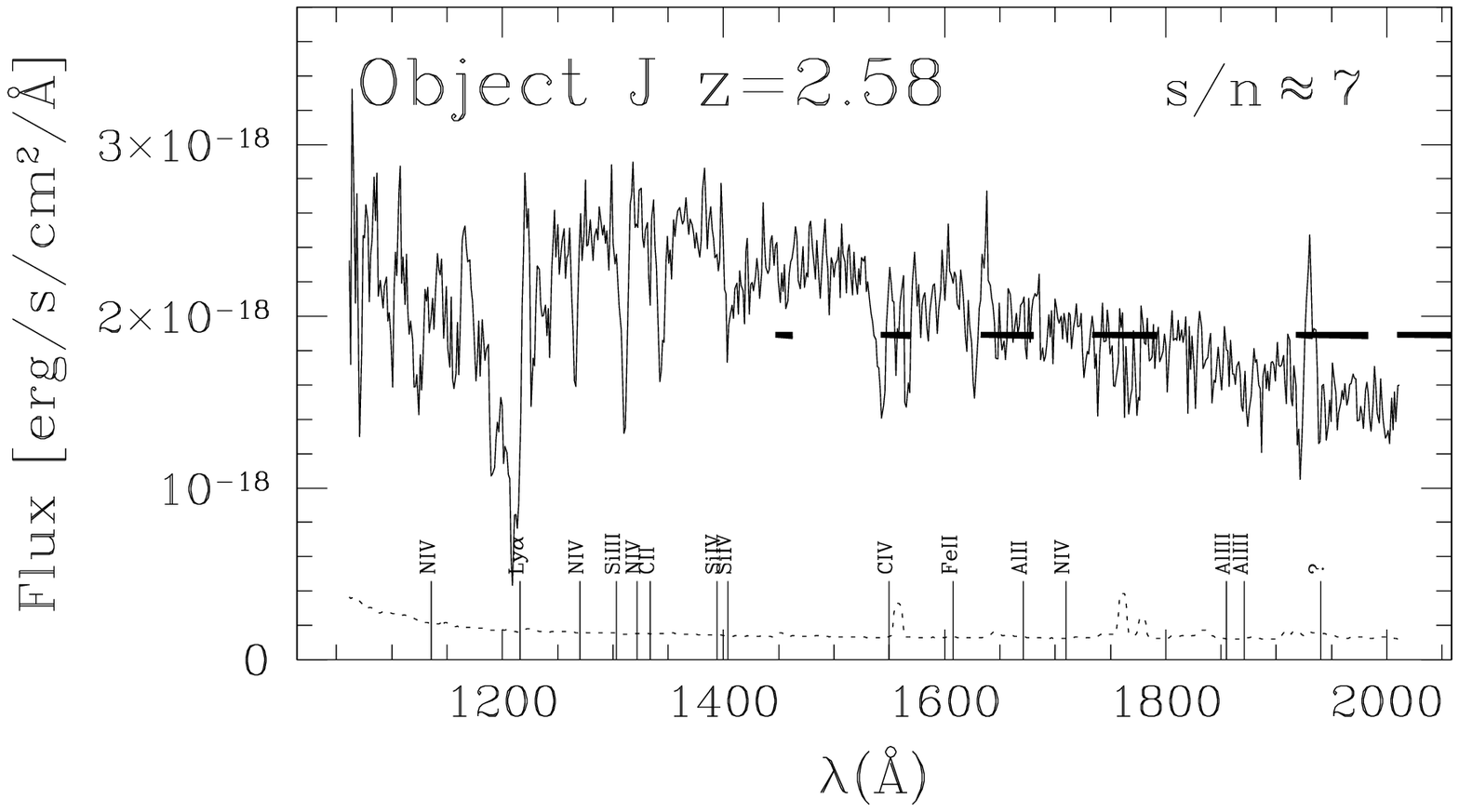,width=10cm}}}
\caption{Spectrum of object~J, which lies $\approx 2.5'$ SW of the arc.
Ly$_{\alpha}$ absorption and metal lines are indicated as in
Figure~\ref{fig:spec_arc}.}
\label{fig:spec_j}
\end{figure}

Most surprisingly even at a distance of $\approx 2.5'$ SW from the arc
we find another high redshift object (J; Figure~\ref{fig:spec_j}),
which either is coincident by chance or points to a large extent of
the total cluster mass.
The latter is not implausible since the subclumps of the cluster
are spread over almost the whole field of view.
Again the Ly$_{\alpha}$ absorption line is the dominant feature
in the spectrum of object~J, giving a redshift of $z = 2.60$.
Object~J also shows Ly$_{\alpha}$ emission, and from the continuum slope
redwards of Ly$_{\alpha}$ we estimate the temperature of its
hottest O~stars to be about $T \approx 25000\,{\rm K}$,
again strongly dependent on the presence of intrinsic UV extinction.

\begin{figure}
\centerline{\hbox{\psfig{figure=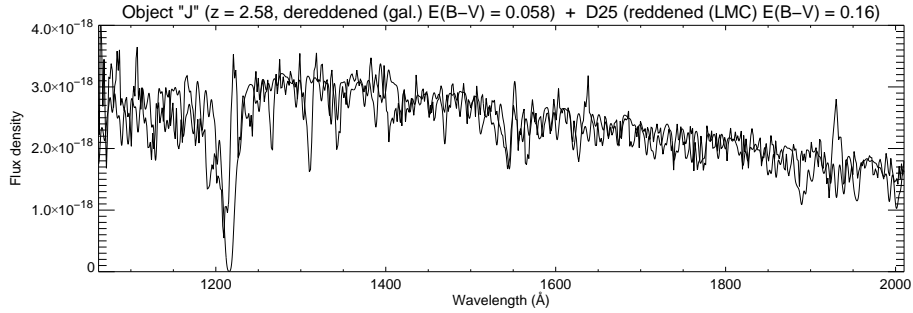,width=12cm,angle=-90}}}
\caption{Comparison of object~J (thin line)
and a synthetic spectrum (thick line).
(See text).}
\label{fig:calc_j}
\end{figure}

As a first attempt to model the observed spectra,
we have calculated the spectrum of a 25000\,K main sequence star
with a metallicity of 1/5 solar
and a mass loss rate of $10^{-9} M_{\sun}/{\rm yr}$,
which Figure~\ref{fig:calc_j} shows overlaid on object~J's spectrum.
To account for both galactic and intrinsic reddening,
we have dereddened the observed spectrum using Howarth's (1983)
analytical expression for galactic extinction with $E(B-V) = 0.058$
(Burstein and Heiles 1984) prior to blueshifting,
and reddened the synthetic spectrum using the LMC extinction law
with $E(B-V) = 0.16$ (the latter value is, of course, a fit parameter).
Apart from Hydrogen
(with a column density of $10^{21}/{\rm cm}^{2}$
at the same assumed redshift as the observation,
i.\,e., the rest frame of the stellar spectrum)
we have not yet added interstellar absorption lines
to the synthetic spectrum.

\begin{table}
\begin{center}
\begin{tabular}{|c|c|c|c|c||c|}
\hline
Object & $R_{\rm obs}$ & s/n & magnification & $R_{\rm i}$ & intr.\ exposure\\
  & (mag) & & (mag) & (mag) & time (s/n $\approx 2.5$)\\
\hline
arc/core & 21.5 & 5/7 & 3      & 24.5       & 10 h\\
A        & 22.7 & 4   & 1--2   & 23.7--24.7 & 2--14 h\\
C        & 23.2 & 3   & 2.5    & 25.7       & 85 h\\
J        & 22.3 & 7   & 0.8--2 & 23.1--24.3 & 1--7 h\\
B        & 23.8 & 3   & 1--2   & 24.8--25.8 & 17--110 h\\
\hline
\end{tabular}
\end{center}
\caption[]
{Observed R-band magnitude;
approximate signal-to-noise ratio of the spectrum obtained within 1.5\,h,
gravitational magnification of the background source,
demagnified R-band magnitude $R_{\rm i}$ and
exposure time that would have been required without
gravitational amplification to reach s/n $\approx 2.5$
(slit width $= 1''$; seeing $= 0.8''$).}
\label{tab:effectivity}
\end{table}

In Table~\ref{tab:effectivity} we list the observed R-band magnitudes
of the objects investigated spectroscopically.
Additionally we list the magnification of the background sources
due to the gravitational lens as well as the R-band magnitudes that
would be observed in the absence of gravitational amplification.
To demonstrate the efficiency of the gravitational telescope
we used the exposure time calculator of FORS provided by ESO and
estimated the total exposure time that would have been necessary to
obtain a spectrum of s/n $\approx 2.5$ from the unmagnified galaxies.
These numbers show that only by having a gravitational telescope at hand
we could obtain acceptable s/n spectra of the distant background sources
visible near 1E\,0657 with a reasonable exposure time.
This confirms that gravitational telescopes offer a unique possibility to
investigate the physics and chemistry of high redshift galaxies in detail.

\section{Future Plans}

With the available spectra we plan to constrain the stellar population
of the observed high redshift galaxies using NLTE hot star models
(Pauldrach et al.\ 1998).
From the line ratios we hope to derive the mean metallicity
of the observed population.

Moreover, the stretched arc itself will be used in an attempt
to investigate the spatial distribution of the stellar population
and the kinematics of this high redshift galaxy.
In principle, reconstructing the original image of the galaxy
by means of a lens model should allow us to investigate its
dynamical state and estimate its total mass.

{\bf Acknowledgements:}
We are grateful to M.\,Ramella for sharing his data and knowledge
about earlier imaging and spectroscopic observations of 1E\,0657,
which helped us in optimizing the observational strategy
and improved the interpretation of the data. This research was supported by 
the German Science Foundation (DFG, SFB 439).
%
%

\end{document}